\documentclass[preprint,5p,times,pdftex]{elsarticle}

\biboptions{sort&compress}

\usepackage{amsmath}
\usepackage{amssymb}
\usepackage{color}
\usepackage{color}
\usepackage{url}
\usepackage{amsfonts}
\usepackage{numprint}

\usepackage[pagebackref=false,backref=false]{hyperref}
\hypersetup{
  pdftitle={A novel adaptive time stepping variant},
  pdfauthor={M. Toggweiler, A. Adelmann, P. Arbenz, J. J. Yang},
  pdfkeywords={}
}

\usepackage{numprint}
\npthousandsep{'}
\npdecimalsign{.}
\npfourdigitnosep

\usepackage{booktabs} 
 \usepackage{algorithm}
\usepackage{algorithmic}
\newcommand{\V}[1]{\mathbf{#1}} 

\newcommand{\opal}{\textsc{OPAL}}

\newcommand{\bb}{Bo\-ris--Bu\-ne\-man}

\newdefinition{remark}{Remark}
\newcommand {\RM}[1]{\mathrm{#1}}

\journal{Journal of Computational Physics (JCP)}

\begin{document} 

\begin{frontmatter}

  \title{A novel adaptive time stepping variant of the \bb\ integrator
    applied to particle accelerator simulation with space charge}
 
  \title{A novel adaptive time stepping variant of the \bb\ integrator\\
    for the simulation of particle accelerators with space charge}

  \author[eth,psi,mit]{Matthias Toggweiler}
  \ead{rmf7@m4t.ch}

  \author[psi]{Andreas Adelmann\corref{cor}}
  \ead{andreas.adelmann@psi.ch}
  
  \author[eth]{Peter Arbenz}
  \ead{arbenz@inf.ethz.ch}

  \author[psi]{Jianjun J. Yang}
  \ead{jianjun.yang@psi.ch}

  \address[eth]{ETH Z\"urich, Computer Science Department,
    Universit\"atsstrasse 6, 8092 Z\"urich, Switzerland}

  \address[psi]{Paul Scherrer Institute, CH-5234 Villigen, Switzerland}

  \address[mit]{MIT, Department of Physics, 77 Massachusetts Avenue, MA 02139, United States}

  \cortext[cor]{Corresponding author}

  \begin{abstract}
    We show that adaptive time stepping in particle accelerator simulation
    is an enhancement for certain problems. The new algorithm has been
    implemented in the \opal\ (Object Oriented Parallel Accelerator Library)
    framework.\
    The idea is to adjust the frequency of costly self field calculations,
    which are needed to model Coulomb interaction (space charge) effects. In
    analogy to a Kepler orbit simulation that requires a higher time step
    resolution at the close encounter, we propose to choose the time step
    based on the magnitude of the space charge forces. Inspired by geometric
    integration techniques, our algorithm chooses the time step proportional
    to a function of the current phase space state instead of calculating a
    local error estimate like a conventional adaptive procedure. 
    Building on recent work, a more profound 
    argument is given on how exactly the time step should be chosen. An
    intermediate algorithm, initially built to allow a clearer analysis by
    introducing separate time steps for external field and self field
    integration, turned out to be useful by its own, for a large class of
    problems.
  \end{abstract}
  \begin{keyword}
    Adaptive time stepping \sep Boris-Buneman \sep particle-in-cell \sep
    space charge \sep particle accelerator simulation
  \end{keyword}
\end{frontmatter}

\section{Introduction}
\label{sec:intro}

In recent years, precise beam dynamics simulations in the design of
high-current low-energy hadron machines as well as of 4th generation
light sources have become a very important research topic.  Hadron
machines are characterized by high currents and hence require excellent
control of beam losses and/or keeping the emittance (a measure of the
phase space) of the beam in narrow ranges. This is a challenging problem
which requires the accurate modeling of the dynamics of a large ensemble
of macro or real particles subject to complicated external focusing,
accelerating and wake-fields, as well as the self-fields caused by the
Coulomb interaction of the particles.

The simulation method discussed in this paper is part of a general accelerator
modeling tool, \opal~(Object Oriented Parallel Accelerator Library)
\cite{opal}.  \opal\ allows to tackle the most challenging problems in
the field of high precision particle accelerator modeling.  These
include the simulation of high power hadron accelerators and of next
generation light sources. Recent physics proposals include~\cite{AgarwallaConradShaevitz,daedalus,isodar}, all of them require large scale
particles based simulation in order to design and optimize the required high power hadron machines.

Here, we discuss methods which track the orbit of each particle
individually, with time as the independent variable.  Accurate modeling
demands the usage of many simulation particles, which of course makes
this approach expensive.
The flow of a particle is
described as an initial value problem for a differential equation,
therefore such methods are called (time) integrators. What is common to
them is that they create a discrete trajectory that approximates the
solution of the initial value problem. How they transport a given state
from time $t_n$ to $t_{n+1}$ is crucial for accuracy and computational
effort.

Currently, only time integrators that use constant time steps $\Delta t
= t_{n+1}\! -\! t_n$ are utilized in \opal. The goal of the recent
master thesis~\cite{ms-thesis} was to investigate whether an integrator
that uses variable time steps can provide enhanced efficiency.  Since
the space charge solver, which computes the self field induced by the
charge, requires all particles to be synchronized in time,
there is a global time step shared among all the particles.  Only this
global time step is to be adapted.  Other kinds of adaptation, like
adaptive mesh refinement, are of separate concern and not the subject of
this work.

In \cite{ms-thesis} two categories of adaptive time stepping schemes were
identified:
\renewcommand{\labelenumi}{(\alph{enumi})}
\begin{enumerate}
\item conventional adaptivity that modifies the time step based on a
  local error estimate, and
\item geometric adaptive integration which aims at solving a regularized
  differential equation.
\end{enumerate}
Both categories adapt the time step based on the local dynamics.
However, the way how this is achieved differs.  In (a), the step
size is changed such that a predicted local error is below some target
value, assuming that the local error is roughly proportional to some
power of the step size.  The local error estimate is usually obtained by
comparing two local solutions of different order, like, e.g., in
Runge-Kutta schemes with step size control~\cite{gear71}.\ Typical for this kind of
algorithms is that a step can be rejected and repeated if the new local
error estimate is above the tolerance. On the other hand, the origin
of (b), geometric adaptive integration, are the deficiencies of (a) when
applied to systems with special structure, e.g.\ Hamiltonian
systems. Integrators, made adaptive in the conventional way, lose their
reversibility or symplecticity properties.

First, it was not clear, how to obtain an appropriate and cheap-to-compute
local error estimate in our particle-in-cell application for approach
(a), and the possibility that steps have to be rejected and repeated
indicated that the memory consumption may increase. The reason why we
considered approach (b) was the hope that it would present a
light-weight alternative with less overhead.  Strict time-reversibility
of the adaptive scheme was not a major issue.

A Kepler orbit problem was chosen as a model problem of adaptive geometric
integration.  A leapfrog method, adaptive in the sense of~(b),
was used to perform smaller steps when the comet is near the sun and
attracted by large forces, and larger steps when the comet is far away.
Following this model, we derived an adaptive variant of the \bb\ method.
Where in the Kepler case we adapted the time step to the strength of the
gravity force, in our application it is adapted to the strength of space
charge forces.  In both cases, the strength of the force is proportional
to the square of the distance, just the sign and the coupling constant
is different for the Coulomb force.

A first implementation proved this type of adaptive time stepping to be
beneficial for particle accelerator simulations, see the results on two
photoinjector scenarios in~\cite{ms-thesis}.  However, a clear argument
was missing on how the time step should be chosen for optimal results.
Moreover, time step requirements for external fields were not
considered.  In this follow-up work, based on a 1D model problem, we
give a sound motivation for the used adaptation scheme. The enhanced version of the adaptive algorithm still chooses the time step based on space charge forces
only, but allows finer inner time steps in a multiple-time-stepping
fashion to account for fast varying external fields.

The initial hope was that we can show the usefulness of variable time
steps for many other scenarios too. However, the benefit over constant
steps with our algorithm becomes visible only when variations in space
charge forces are large. This is the case for space charge-dominated
scenarios like a photoinjector simulation, but e.g.\ not for simulations
of high-energy cyclotron turns. For cases where variations in the space
charge forces are small but space charge still has a visible effect, the
multiple-time-stepping extension alone proved to be an important
enhancement to the original \bb\ method. Although it was thought to be
only an intermediate algorithm and its derivation was relatively easy
compared to the rest, this part is probably the achievement with broader
practical impact.

In Section~\ref{sec:SCsims} we state the problem.  In
Section~\ref{sec:constant} follows an introduction to the current time
integration scheme available in \opal. The new adaptive \bb\ scheme is
presented in Section~\ref{sec:adapt}, building on an
multiple-time-stepping extension of the original method.  In
Section~\ref{sec:1D-model} we examine more closely the adaptive step
size strategy by looking at a one-dimensional model problem. In
Section~\ref{sec:applications} we benchmark the developed methods on
applications from beam dynamics, and in Section~\ref{sec:conclusions} we
draw our conclusions.


\section{Particle beam simulations with space charge}
\label{sec:SCsims}

\subsection{The electrostatic particle-in-cell method}

We consider the
Vlasov-Poisson description of the phase space, including external and
self fields.  Let $f(\mathbf{x},\mathbf{v},t)$ be the density of the particles in the
phase space, i.e.\ the position-velocity $(\mathbf{x}, \mathbf{v})$
space.  Its evolution is determined by the collisionless \emph{Vlasov
  equation},
\begin{equation*} \label{eq:Vlasov}
  \frac{df}{dt}=\partial_t f + \mathbf{v} \cdot \nabla_{\mathbf{x}} f
  +\frac{q}{m}(\mathbf{e}+ \mathbf{v}\times\mathbf{b})\cdot
  \nabla_{\mathbf{v}} f  =  0,
\end{equation*}
where $m$, $q$ denote particle mass and charge, respectively.  The
electric and magnetic fields $\mathbf{e}$ and $\mathbf{b}$ are
superpositions of external fields and self fields (space charge),
\begin{equation}\label{eq:allfield}
    \mathbf{e} =
    \mathbf{e^{\RM{ext}}} + \mathbf{e^{\RM{self}}}, \qquad
    \mathbf{b} =
    \mathbf{b^{\RM{ext}}} + \mathbf{b^{\RM{self}}}.
\end{equation}
If $\mathbf{e}$ and $\mathbf{b}$ are known, then each particle can be
propagated according to the equation of motion for charged particles in an
electromagnetic field. After particles have moved, we have to 
update $\mathbf{e^{\RM{self}}}$ and $\mathbf{b^{\RM{self}}}$ (among other things).  
For this, we change the coordinate system into the one moving with the
particles.  By means of the appropriate \emph{Lorentz
  transformation $\mathcal{L}$}~\cite{lali:84} we arrive at a (quasi-) static
approximation of the system in which the transformed magnetic field
becomes negligible, $\hat{\mathbf{b}}\! \approx\! \mathbf{0}$.  The
transformed electric field is then obtained from
\begin{equation}\label{eq:e-field}
  \hat{\mathbf{e}}=\hat{\mathbf{e}}^{\RM{self}}=-\nabla\hat{\phi},
\end{equation}
where the electrostatic potential $\hat{\phi}$ is the solution of the
\emph{Poisson problem}
\begin{equation}\label{eq:poisson0}
  - \Delta \hat{\phi}(\mathbf{x}) =
  \frac{ \mathcal{L}(\rho(\mathbf{x}))}{\varepsilon_0},
\end{equation}
equipped with appropriate boundary conditions.  Here, $\rho$ denotes the spatial charge
density and $\varepsilon_0$ is the dielectric constant.
By means of the inverse Lorentz transformation ( $\mathcal{L}^{-1}$) the electric field
$\hat{\mathbf{e}}$ can then be transformed back to yield both the
electric and the magnetic fields in~\eqref{eq:allfield}.

The Poisson problem~\eqref{eq:poisson0} discretized by finite
differences can efficiently be solved on a rectangular grid by a
Particle-In-Cell (PIC) approach~\cite{qiry:01}.  The right hand side
of~\eqref{eq:poisson0} is discretized by sampling the particles at the
grid points.  In~\eqref{eq:e-field}, $\hat{\phi}$ is interpolated at the
particle positions from its values at the grid points. We also note that
the FFT-based Poisson solvers and similar
approaches~\cite{qiry:01,qigl:04} are most effective in box-shaped or open domains.

\subsection{Equations of motion}

We integrate in time $N$ identical particles, all having the rest mass $m$
and charge $q$.  The relativistic equations of motion for particle $i$
are
\begin{align}
  \frac{\mathrm{d}\V{x}_i}{\mathrm{d}t} &= \frac{\V{p}_i}{m \gamma_i}, \\
\frac{\mathrm{d}\V{p}_i}{\mathrm{d}t} &= q \left(\V{e}_i+\frac{\V{p}_i}{m \gamma_i} \times \V{b}_i\right), \label{eq:dpdt}
\end{align}
where $\V{x}_i$ is the position, $\V{p}_i = m \V{v}_i \gamma_i$ the relativistic momentum, $\V{v}_i$ the velocity,
$\gamma_i = 1 / \sqrt{1-(||\V{v}_i||/c)^2} = \sqrt{1+(||\V{p}_i||/(mc))^2}$ the Lorentz factor and $c$ the speed of light. The electric and
magnetic field, $\V{e}_i$ and $\V{b}_i$, can be decomposed into external field and self field contributions:
\begin{align}
\V{e}_i &= \V{e}^{\mathrm{ext}}(\V{x}_i, t) + \V{e}^{\mathrm{self}}(i, \V{x}_{1 \ldots N}, \V{p}_{1 \ldots N}), \\
\V{b}_i &= \V{b}^{\mathrm{ext}}(\V{x}_i, t) + \V{b}^{\mathrm{self}}(i, \V{x}_{1 \ldots N}, \V{p}_{1 \ldots N}).
\end{align}
The notation $\V{x}_{1 \ldots N}$ is a shorthand for $\V{x}_1, \ldots,
\V{x}_N$, and is used for other vectors analogously. The self field describes the
field created by the collection of particles i.e. the source of the Coulomb repulsion.
The external electromagnetic field (from magnets etc.), which can have an explicit
dependence on time $t$, are in this model treated independent of the other particles.


\section{Constant time step integration method}
\label{sec:constant}

The \bb\ integration scheme \cite[pp.~58--63]{plasmaPhysics} sol\-ves the
single particle equations of motion in electric and magnetic fields, and
is a widely used orbit integrator in explicit PIC simulations and codes
such as \opal. The scheme is popular because it is simple to
implement and because it gives second order accuracy while requiring 
only one force evaluation per step.

The integration method is similar to the leapfrog method (see e.g.\ \cite{ms-thesis}) 
as it offsets position and momentum by half a time step and updates them
alternatingly. However, we have to deal with a non-separable
system as the momentum derivative depends on the momentum
itself in the magnetic force term. This requires special
treatment to retain good properties like time-symmetry.

The classic derivation by Buneman and Boris is explained in \cite{ms-thesis}
for the nonrelativistic case. The relativistic generalization is not difficult 
\cite[pp.~356--357]{plasmaPhysics}. Here, we develop the method by an operator
splitting approach \cite[{sec.\ II.5}]{geomNumInt}, directly working on the relativistic 
equations of motion. We will see that this yields exactly the same method. 
While this alternative derivation itself is not strictly necessary for this paper, it serves 
to introduce the operator-notation used to characterize the new integrator. 
In the operator-notation we make a simplifying assumption about the fields and
look at only one particle, in order to allow the reader to quickly grasp the ideas. In the algorithms we will give all details how 
the integrators are implemented in our specific application.

We now look at a single particle and assume that the fields
depend only on position $\V{x}$, i.e.\ do not depend on $\V{p}$ and are
constant over time. Let the equations of motion be written as
\begin{equation} \label{eq:dpdt_vec}
  \frac{\mathrm{d}}{\mathrm{d}t}
  \left(\begin{array}{c}
      \V{x}\\ 
      \V{p}\end{array}\right)=
  \left(\begin{array}{c}
      \V{f}_X(\V{p})\\ 
      \V{0}\end{array}\right) +
  \left(\begin{array}{c}
      \V{0}\\ 
      \V{f}_E(\V{x})\end{array}\right) +
  \left(\begin{array}{c}
      \V{0}\\ 
      \V{f}_B(\V{x},\V{p})\end{array}\right),
\end{equation}
where
\begin{align*}
  \V{f}_X(\V{p}) &= \V{p}/(m\gamma(\V{p})), \\
  \V{f}_E(\V{x}) &= q\V{e}(x), \\
  \V{f}_B(\V{x},\V{p}) &= q\V{p} \times \V{b}(\V{x})/(m\gamma({\V{p})}).
\end{align*}
For simplified systems where only one term of the right hand side of~\eqref{eq:dpdt_vec}
exists we can make the following statements:
\begin{itemize}
\item If only $\V{f}_X$ was present in the RHS, then
\begin{align*}
\text{Drift}_h: \left(\begin{array}{c}
\V{x}\\ 
\V{p}\end{array}\right) \mapsto \left(\begin{array}{c}
\V{x} + h\V{f}_X(\V{p})\\ 
\V{p}\end{array}\right)
\end{align*}
would be the flow of the system, i.e.\ would integrate the system exactly.
\item If only $\V{f}_E$ was present in the RHS, then
  \begin{align*}
    \text{E-Kick}_h: \left(\begin{array}{c}
        \V{x}\\ 
        \V{p}\end{array}\right) \mapsto \left(\begin{array}{c}
        \V{x}\\ 
        \V{p}+h\V{f}_E(\V{x})\end{array}\right)
  \end{align*}
  would be the flow of the system.
\item If only $\V{f}_B$ was present in the RHS, then
  \begin{align*}
    \text{B-Kick}_h: \left(\begin{array}{c}
        \V{x}\\ 
        \V{p}\end{array}\right) \mapsto \left(\begin{array}{c}
        \V{x}\\ 
        \V{p}+h\V{f}_B(\V{x},\V{p})\end{array}\right)
  \end{align*}
  would be a numerical first-order approximation to the flow of the system
  (forward Euler method).
\end{itemize}
Having identified these parts, we construct a second-order integrator by
using a suitably ordered composition of these mappings.  By two nested
  applications of equation (5.9)
  \cite[p. 49]{geomNumInt} (Combining Exact and Numerical Flows),
\begin{displaymath}
  \Phi_h =    \text{B-Kick}_h \circ \text{E-Kick}_h  \circ  \text{Drift}_h 
\end{displaymath}
is a first-order integrator for the combined system. 
{By composing the operator $\Phi_{h/2}$ with its adjoint
  $\Phi_{h/2}^*$ a second order integrator is obtained~\cite[p.~45]{geomNumInt},}
\begin{align*}
  \text{BB}_h &= \Phi_{h/2}^* \circ \Phi_{h/2} = (\Phi_{-h/2})^{-1}
  \circ \Phi_{h/2}  \\
  &=  \text{Drift}_{h/2}~ \circ \nonumber\\
  & \quad \underbrace{\text{E-Kick}_{h/2} \circ
    (\text{B-Kick}_{-h/2})^{-1} \circ \text{B-Kick}_{h/2} \circ
    \text{E-Kick}_{h/2}}_{\displaystyle\text{Kick}_h} \nonumber\\
  &\quad \circ \text{Drift}_{h/2},\nonumber
\end{align*}
where the second equality recalls the definition of the adjoint. This is already the \bb-method.
However, the magnetic field kick in the middle
\begin{displaymath}
\left(\begin{array}{c}
\V{x}_{**}\\ 
\V{p_{**}}\end{array}\right) = (\text{B-Kick}_{-h/2})^{-1} \circ \text{B-Kick}_{h/2}
\left(\begin{array}{c}
\V{x}_{*}\\ 
\V{p_{*}}\end{array}\right)
\end{displaymath}
is an implicit mapping. The $\V{p}$-component is 
\begin{displaymath}
  \begin{split}
    \V{p}_{**} &= \V{p}_{*} + hq\V{p}_{*} \times
    \V{b}(\V{x}_{*})/(2m\gamma(\V{p}_{*})) + \\ 
    &\qquad hq\V{p}_{**} \times \V{b}(\V{x}_{**})/(2m\gamma(\V{p}_{**})),
  \end{split}
\end{displaymath}
which is recognizable as the trapezoidal integration rule.
Using that $\V{x}_*=\V{x}_{**}$ and that the length of the momentum vector is
invariant in the B-field rotation ($\gamma$ stays constant), we have
\begin{displaymath}
\frac{\V{p}_{**} - \V{p}_{*}}{h}=q \frac{\V{p}_{*} + \V{p}_{**}}{2m\gamma(\V{p}_{*})}
 \times \V{b}(\V{x}_{*}).
\end{displaymath}
Boris' important contribution is the way how $\V{p}_{**}$ can be calculated explicitly,
thus making the whole method explicit.

In our application, the fields do not only depend on the position. But
the change with respect to momentum and time are typically small within
a time step, so the properties of the method still hold approximatively
(assuming the time steps are not chosen too large).  The method is
implemented in \opal\ as described in Algorithms 1, 2, and 3.

\begin{algorithm}
\caption{BB($\V{x}_{1 \ldots N}, \V{p}_{1 \ldots N}, t_{\text{end}}, h$)}
\label{BB-algo}
\begin{algorithmic}
\STATE $t \leftarrow 0$
\WHILE{$t < t_{\text{end}}$}
\STATE $(\V{x}_{1 \ldots N}, t) \leftarrow \text{Drift}\left(h/2, \V{x}_{1 \ldots N}, \V{p}_{1 \ldots N},t \right)$
\STATE $\left(\V{e}_{1 \ldots N}, \V{b}_{1 \ldots N}\right) \leftarrow \text{ExternalFields}\left(\V{x}_{1 \ldots N}, t\right) +$ \\ \hspace{0.5cm} $\text{SelfField}\left(\V{x}_{1 \ldots N}, \V{p}_{1 \ldots N}\right)$
\STATE $\V{p}_{1 \ldots N} \leftarrow \text{Kick}\left(h, \V{p}_{1 \ldots N}, \V{e}_{1 \ldots N}, \V{b}_{1 \ldots N}\right)$
\STATE $(\V{x}_{1 \ldots N}, t) \leftarrow \text{Drift}\left(h/2, \V{x}_{1 \ldots N}, \V{p}_{1 \ldots N},t \right)$
\ENDWHILE
\end{algorithmic}
\end{algorithm}

\begin{algorithm}
\caption{Drift$\left(h, \V{x}_{1 \ldots N}, \V{p}_{1 \ldots N}, t  \right)$}
\begin{algorithmic}
\FOR{$i=1$ to $N$}
\STATE $\gamma \leftarrow \sqrt{1+\V{p}_i^\text{T}\V{p}_i/(mc)^2}$
\STATE $\V{x}_i \leftarrow \V{x}_i + h\V{p}_i/(m \gamma)$
\ENDFOR
\STATE $t \leftarrow t + h$
\RETURN $\left(\V{x}_{1 \ldots N}, t \right)$
\end{algorithmic}
\end{algorithm}

\begin{algorithm}
\caption{Kick$\left(h, \V{p}_{1 \ldots N}, \V{e}_{1 \ldots N}, \V{b}_{1 \ldots N} \right)$}
\begin{algorithmic}
\FOR{$i=1$ to $N$}
\STATE $\V{p}_i \leftarrow \V{p}_i + hq\V{e}_i/2$
\STATE $\gamma \leftarrow \sqrt{1+\V{p}_i^\text{T}\V{p}_i/(mc)^2}$
\STATE $\V{r} \leftarrow hq\V{b}_i/(2m\gamma)$ \\
\STATE $\V{w} \leftarrow \V{p}_i + \V{p}_i \times \V{r}$ \\
\STATE $\V{s} \leftarrow 2\V{r}/(1+\V{r}^\text{T} \V{r})$ \\
\STATE $\V{p}_i \leftarrow \V{p}_i + \V{w} \times \V{s}$ \\
\STATE $\V{p}_i \leftarrow \V{p}_i + hq\V{e}_i/2$
\ENDFOR
\RETURN $\V{p}_{1 \ldots N}$
\end{algorithmic}
\end{algorithm}



\section{Adaptive time step integration method}
\label{sec:adapt}

\subsection{Multiple-time-stepping (MTS) variant} \label{subsec:mts}
Before discussing our adaptive step size variant of the \bb\ integrator,
we discuss a simple but powerful extension to the fixed time integration
scheme from chapter 3.  We are following~\cite[Chapter VIII
4.1]{geomNumInt} and write the differential equation with a fast-slow
splitting as
\begin{align}
\frac{\mathrm{d}\V{p}}{\mathrm{d}t} &= \V{f}^{\mathrm{ext}} + \V{f}^{\mathrm{self}}.
\end{align}
$\V{f}^{\mathrm{self}}$ corresponds to the slow dynamics and is also the
most expensive term to evaluate, namely the space charge forces. The
fast dynamics is governed by $\V{f}^{\mathrm{ext}}$, and in our
particular example, this term arises from fast varying external fields
in an accelerator. For details see Algorithm~4.1 and Lemma~4.2 of
\cite[{sec.\ VIII.4}]{geomNumInt}. We state a multiple-time-stepping variant of the \bb\
integrator as
\begin{align}
\label{eq:MTS}
\mathrm{MTS}_{h}^{m} = \mathrm{Kick}_{h/2}^{\mathrm{self}} \circ \left(\mathrm{BB}_{h/m}^{\mathrm{ext}}\right)^m \circ \mathrm{Kick}_{h/2}^{\mathrm{self}}
\end{align}
where the $m \geq 1$ substeps are defined by
\begin{align}
\mathrm{BB}_{h}^{\mathrm{ext}} = \mathrm{Drift}_{h/2} \circ \mathrm{Kick}_{h}^{\mathrm{ext}} \circ \mathrm{Drift}_{h/2}.
\end{align}
The superscripts ``ext'' and ``self'' of the Kick-operator denote that
only external and self-field forces, respectively, are used for the
momentum update. The pseudo-code of the method is shown in Algorithm
\ref{alg:mts}, which refers to Algorithm~\ref{alg:bbext} and to already
presented Algorithms 1, 2 and 3.

\begin{algorithm}
\caption{MTS($\V{x}_{1 \ldots N}, \V{p}_{1 \ldots N}, t_{\text{end}}, h, m$)}
\label{alg:mts}
\begin{algorithmic}
\STATE $\left(\V{e}^{\text{self}}_{1 \ldots N}, \V{b}^{\text{self}}_{1 \ldots N}\right) \leftarrow \text{SelfField}\left(\V{x}_{1 \ldots N}, \V{p}_{1 \ldots N}\right)$
\STATE $t \leftarrow 0$
\WHILE{$t < t_{\text{end}}$}
\STATE $\V{p}_{1 \ldots N} \leftarrow \text{Kick}\left(h/2, \V{p}_{1 \ldots N}, \V{e}^{\text{self}}_{1 \ldots N}, \V{b}^{\text{self}}_{1 \ldots N}\right)$
\FOR{$i = 1, \ldots, m$}
\STATE $\left(\V{x}_{1 \ldots N}, \V{p}_{1 \ldots N}, t \right) \leftarrow \mathrm{BB}^{\mathrm{ext}}\left(h/m, \V{x}_{1 \ldots N}, \V{p}_{1 \ldots N}, t\right)$
\ENDFOR
\STATE $\left(\V{e}^{\text{self}}_{1 \ldots N}, \V{b}^{\text{self}}_{1 \ldots N}\right) \leftarrow \text{SelfField}\left(\V{x}_{1 \ldots N}, \V{p}_{1 \ldots N}\right)$
\STATE $\V{p}_{1 \ldots N} \leftarrow \text{Kick}\left(h/2, \V{p}_{1 \ldots N}, \V{e}^{\text{self}}_{1 \ldots N}, \V{b}^{\text{self}}_{1 \ldots N}\right)$
\ENDWHILE
\end{algorithmic}
\end{algorithm}

\begin{algorithm}
\caption{$\mathrm{BB}^{\mathrm{ext}}$($h, \V{x}_{1 \ldots N}, \V{p}_{1 \ldots N}, t$)}
\label{alg:bbext}
\begin{algorithmic}
\STATE $(\V{x}_{1 \ldots N}, t) \leftarrow \text{Drift}\left(h/2, \V{x}_{1 \ldots N}, \V{p}_{1 \ldots N},t \right)$
\STATE $\left(\V{e}^{\text{ext}}_{1 \ldots N}, \V{b}^{\text{ext}}_{1 \ldots N}\right) \leftarrow \text{ExternalFields}\left(\V{x}_{1 \ldots N}, t\right)$
\STATE $\V{p}_{1 \ldots N} \leftarrow \text{Kick}\left(h, \V{p}_{1 \ldots N}, \V{e}^{\text{ext}}_{1 \ldots N}, \V{b}^{\text{ext}}_{1 \ldots N}\right)$
\STATE $(\V{x}_{1 \ldots N}, t) \leftarrow \text{Drift}\left(h/2, \V{x}_{1 \ldots N}, \V{p}_{1 \ldots N},t \right)$
\RETURN $\left(\V{x}_{1 \ldots N}, \V{p}_{1 \ldots N}, t \right)$
\end{algorithmic}
\end{algorithm}



\subsection{Adaptive step size variant}

We explained in the introduction why geometric adaptive integration was
used as the starting point in the design of our algorithm. Briefly, the simplicity of choosing the time
step proportional to the system's state without having to deal with
error estimates and step rejection was compelling, and the results of
the previous work~\cite{ms-thesis} motivated to further investigate this
approach.
 
The concept of adaptive geometric integration is to look at a transformed system that evolves in a fictitious time $\tau$. The relation that connects the real time to it is called a Sundman transformation
\begin{align}
\frac{\mathrm{d}t}{\mathrm{d}\tau} = g(\V{z}).
\end{align}
The function $g$ determines the time rescaling and depends on the state of the system. It has to be chosen problem-dependent. The idea is that the rescaled system can be integrated with constant steps because the dynamics are regularized. When $g$ is large, $t(\tau)$ increases rapidly and we take large steps in real time. When $g$ is small, $t(\tau)$ increases slowly and we take small steps. 
Where the evolution of the system over $t$ was described with
\begin{align}
\frac{\mathrm{d}\V{z}}{\mathrm{d}t} = \V{f}(\V{z}, t),
\end{align}
the transformed equation with independent variable $\tau$ becomes, by the chain rule of differentiation,
\begin{align}
\frac{\mathrm{d}\V{z}}{\mathrm{d}\tau} = \frac{\mathrm{d}\V{z}}{\mathrm{d}t} \frac{\mathrm{d}t}{\mathrm{d}\tau} = \V{f}(\V{z}, t) g(\V{z}).
\end{align}
The curve $\V{z}(\tau)$ is identical to $\V{z}(t)$, but the trajectories
are traversed at different velocities with respect to the independent
variable. Constant steps taken in the fictitious time $\tau$ correspond
to variable steps in the real time $t$, excluding the uninteresting case
of a constant $g$. Multiple methods exist to solve such a transformed
system.

Preserving time symmetry in the construction of the adaptive integration scheme is an important point for some applications, especially in 
few body settings like the Kepler problem. In~\cite{ms-thesis} we noted that this symmetry is not important for our PIC application, as a consequence  
we do not put effort in enforcing this symmetry, favoring simplicity. The core concept we take over from the adaptive geometric integration 
technique is the choice of the time step proportional to some -- cheap to evaluate -- function of the current state.

How should we choose the function $g$ for our PIC application? Our inspiration came from the treatment of the Kepler problem. There the adaptive 
geometric integration technique demands smaller step sizes in situations where the force is large, i.e.\ when the two bodies are close to each other. In our case, we similarly
want to demand smaller step sizes when particles are near to each other, i.e.\ when the beam volume is small and the space charge forces are large. Of course there
is still a variety of ways how this can be done, especially because we have not only two bodies but a collection of particles. The key point here is: this approach
to step size adaption is based solely on the strength of the space charge force, and not on the properties of the external fields. While this sounds limiting, the regularization of space charge
solve frequency was the most evident opportunity to save computation cost in our application. It is certainly possible to extend the method with some sort of adaptivity for 
external fields, if required.

In order to determine $g$,  for every particle, we compute its acceleration due to the space charge field with
\begin{align}
\label{acceleration}
\V{a}_i = \frac{\text{d}^2\V{x}_i}{\text{d}t^2} = \frac{1}{m \gamma_i} \left( \V{f}_i^{\text{self}} - \V{p}_i \frac{1}{m^2 c^2 \gamma_i^2} \V{p}_i^{\text{T}} \V{f}_i^{\text{self}} \right),
\end{align}
where $\V{f}_i^{\text{self}}$ is the right hand side of \eqref{eq:dpdt} using only the self field contribution.
The acceleration of largest magnitude among all particles is now related to the value of $g$ as
\begin{align}
g\left(\V{p}_{1 \ldots N}, \V{f}_{1 \ldots N}^{\text{self}}\right) \propto \left(\max_{i} ||\V{a}_i||\right)^{-\beta/2}.
\end{align}
Because only proportionality is important, one can leave out constants
in the calculation of $g$. Section~\ref{sec:1D-model} will give an
argument why $\beta=1.0$ is a good choice by analyzing adaptive
integration of a one-dimensional model problem. In our experiments the
overhead for calculating $g$ was negligible compared to the total time
spent, but one can easily make this cheaper if required, e.g.\ by
dropping the second term of the parenthesized part in
\eqref{acceleration} or by using a mean $\gamma$, avoiding per-particle
square root operations.

In~\cite{ms-thesis}, we used a simpler
version of $g$ which was based on the beam size directly. Also, there
was only one time step. Here, we combine separate time steps for
external and self fields together with variable step sizes for self
fields. The MTS algorithm (see Alg.~\ref{alg:mts}) is made adaptive
(``AMTS'') by
\begin{itemize}
\item using a variable outer time step $h$ adapted to space charge
  strength, and,
\item using a variable $m \in \mathbb{N}$ such that the inner time step
  $h/m$ for external field integration is kept roughly at same size if
  possible.
\end{itemize}
The inner time step, of course, can never be higher than the outer time
step.  But as soon the outer time step exceeds some value, $m$ is
increased to keep the external field step roughly at its original
size. See Algorithm~\ref{alg:AMTS} for the full method. The initial step
sizes must be specified as input.

\begin{algorithm}
\caption{AMTS($\V{x}_{1 \ldots N}, \V{p}_{1 \ldots N}, t_{\text{end}}, \Delta t_{\text{outer}}^{\text{init}}, \Delta t_{\text{inner}}$)}
\label{alg:AMTS}
\begin{algorithmic}
\STATE $\left(\V{e}^{\text{self}}_{1 \ldots N}, \V{b}^{\text{self}}_{1 \ldots N}\right) \leftarrow \text{SelfField}\left(\V{x}_{1 \ldots N}, \V{p}_{1 \ldots N}\right)$
\STATE $\lambda \leftarrow g\left(\V{p}_{1 \ldots N}, \V{e}^{\text{self}}_{1 \ldots N}, \V{b}^{\text{self}}_{1 \ldots N}\right)$
\STATE $\Delta \tau \leftarrow \Delta t_{\text{outer}}^{\text{init}} / \lambda$
\STATE $t \leftarrow 0$
\WHILE{$t < t_{\text{end}}$}
\STATE $\lambda \leftarrow g\left(\V{p}_{1 \ldots N}, \V{e}^{\text{self}}_{1 \ldots N}, \V{b}^{\text{self}}_{1 \ldots N}\right)$
\STATE $h \leftarrow \lambda \cdot \Delta \tau$
\STATE $m \leftarrow [h/\Delta t_{\text{inner}}]$
\STATE $\V{p}_{1 \ldots N} \leftarrow \text{Kick}\left(h/2, \V{p}_{1 \ldots N}, \V{e}^{\text{self}}_{1 \ldots N}, \V{b}^{\text{self}}_{1 \ldots N}\right)$
\FOR{$i = 1, \ldots, m$}
\STATE $\left(\V{x}_{1 \ldots N}, \V{p}_{1 \ldots N}, t \right) \leftarrow \mathrm{BB}^{\mathrm{ext}}\left(h/m, \V{x}_{1 \ldots N}, \V{p}_{1 \ldots N}, t\right)$
\ENDFOR
\STATE $\left(\V{e}^{\text{self}}_{1 \ldots N}, \V{b}^{\text{self}}_{1 \ldots N}\right) \leftarrow \text{SelfField}\left(\V{x}_{1 \ldots N}, \V{p}_{1 \ldots N}\right)$
\STATE $\V{p}_{1 \ldots N} \leftarrow \text{Kick}\left(h/2, \V{p}_{1 \ldots N}, \V{e}^{\text{self}}_{1 \ldots N}, \V{b}^{\text{self}}_{1 \ldots N}\right)$
\ENDWHILE
\end{algorithmic}
\end{algorithm}


\section{Inverse square force integration}
\label{sec:1D-model}

\subsection{Model problem}
\begin{figure}[htb]
  \includegraphics[width=.95\columnwidth]{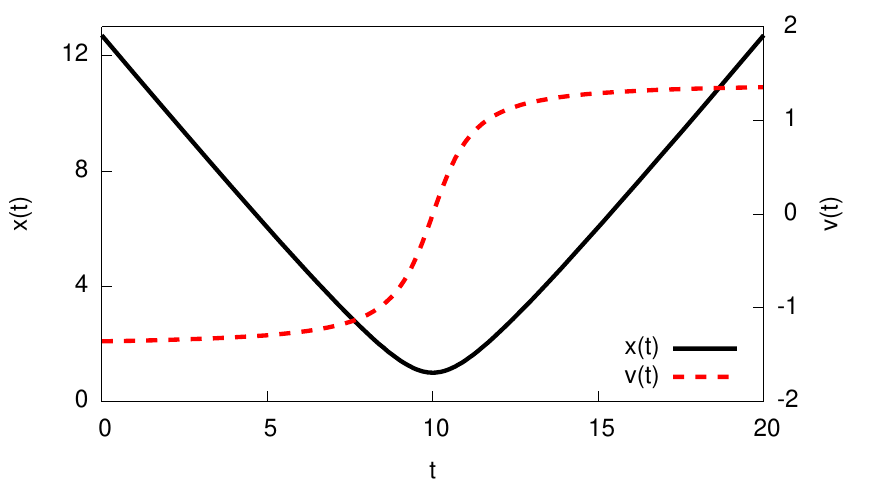}
  \caption{Solution of the 1D hard soft-wall collision problem. Shown are position $x(t)$ and velocity $v(t)$ as function of time.}
  \label{inverseSquareForceSolution}
\end{figure}

To better understand which adaptive strategy is optimal, a simple model problem is considered.
We look at the motion of a single particle in one dimension, under the influence of a repulsive, Coulomb like force.
For position $x$ and velocity $v$ of a single body, let the equations of motion be
\begin{align}
  \frac{\mathrm{d}x}{\mathrm{d}t} &= v, \label{isf-dxdt} \\
  \frac{\mathrm{d}v}{\mathrm{d}t} &= \frac{1}{x^2} \label{isf-dvdt}.
\end{align}
The force is proportional to the inverse square of the distance to the
origin.  Figure~\ref{inverseSquareForceSolution} shows the solution for
the position and velocity as function of time.
We can already `guess' that a smaller time step near the minimum of $x$ (closest encounter) should be used. 
But we don't know how exactly the step size should depend on the situation. Therefore we investigate how the
choice of the time rescaling function $g$ (from some class of candidates) influences the error of adaptive integration methods. The accuracy is
measured by comparing to an analytical solution.

\subsection{Analytical solution}
This specific case allows an analytic treatment. The solution will not be explicit, but can be calculated by finding a root of some function.

Given the initial conditions $x_0 > 0$ and $v_0$, we determine
$x_{\text{L}} = x(t_{\text{L}})$ as the minimum of the curve
$x(t)$.  The energy $H(x,v) = v^2/2 + 1/x$ stays constant along the
solution, and knowing $v(t_{\text{L}}) = 0$ leads to
\begin{displaymath}
  x_{\text{L}} = \frac{2 x_0}{2 + v_0^2 x_0}.
\end{displaymath}
We can give the velocity (only positive solution used here) for any $x \geq x_{\text{L}}$ by using the energy argument again:
\begin{displaymath}
  V(x) = \sqrt{2 \left(\frac{1}{x_{\text{L}}} - \frac{1}{x} \right)}.
\end{displaymath}
Rearranging \eqref{isf-dvdt} using $\mathrm{d}v/\mathrm{d}t =
\mathrm{d}v/\mathrm{d}x \cdot \mathrm{d}x/\mathrm{d}t =
\mathrm{d}v/\mathrm{d}x \cdot v$ lets us integrate on both sides:
\begin{displaymath}
  \int_{x_{\text{L}}}^{x}{\frac{\mathrm{d}V(x)}{\mathrm{d}x} x^2\
    \mathrm{d}x} = \int_0^{T(x)} \mathrm{d}t. 
\end{displaymath}
Evaluating and simplifying gives us the (positive) time for the motion
from $x_{\text{L}}$ up to $x$ with
\begin{displaymath}
  T(x) = \sqrt{\frac{x_{\text{L}}}{2}} \left(\sqrt{x(x-x_{\text{L}})} +
    x_{\text{L}} \log \left(
      \frac{\sqrt{x-x_{\text{L}}}+\sqrt{x}}{\sqrt{x_{\text{L}}}}
    \right)\right).
\end{displaymath}
The final solution is found in two steps. First, we compute the relative
time needed to find the minimal $x$,
\begin{align*}
  t_{\text{L}} =
  \begin{cases}
    -T(x_{\text{L}}), & \text{if $v_0 > 0$,} \\
    T(x_{\text{L}}),  & \text{otherwise,}
  \end{cases}
\end{align*}
then we compute the final condition from
\begin{align*}
  x(t) &=
  \begin{cases}
    T^{-1}(t - t_{\text{L}}), & \text{if $t > t_{\text{L}}$,} \\
    T^{-1}(t_{\text{L}} - t),  &  \text{otherwise,}
  \end{cases} \\
  v(t) &=
  \begin{cases}
    V(x(t)), & \text{if $t > t_{\text{L}}$,} \\
    -V(x(t)),  & \text{otherwise.}
  \end{cases}
\end{align*}
To compute $T^{-1}\left(t\right)$, we find the root of $T(x) - t$ by 
bisecting in the interval $\left[x_{\text{L}},
  x_{\text{L}} + t^2/x_{\text{L}}^2\right]$.

\subsection{Experiment and findings}
We already know that the numerical integration efficiency can be
increased by concentrating effort on parts where the force is large, see
e.g.\ the Kepler problem in~\cite{ms-thesis}. The considered adaptive
methods employ a Sundman transformation of the form
\begin{displaymath}
  \frac{\mathrm{d}t}{\mathrm{d}\tau} = g(x) = x^{\beta} = \left(\frac{\text{d}^2 x}{\text{d}t^2}\right)^{-\beta/2}
\end{displaymath}
applied to the system \eqref{isf-dxdt} -- \eqref{isf-dvdt}. In other words, we aim at making the variable timestep proportional to $g$. Of course
there are other possibilities how to choose the transformation, but the dependeny on solely $x$ means that the timestep will be a function
of the magnitude of the acceleration/force. For problems with many bodies (e.g.\ our PIC simulation), we then can use
the maximal acceleration/force for the choice of a global timestep. The goal is now to
look more closely at the exponent $\beta$ and ask for an optimal value. 

In the following, we integrate three different scenarios until $t_{\text{end}} = 20$.  The
initial conditions are chosen such that the lowest point is reached at
$t_{\text{L}} = 10$, i.e.\ $x_0 = T^{-1}(10)$ and $v_0 = -V(x_0)$. The
scenarios differ in the energy level $H$. A higher energy level means
the magnitude of the initial velocity is larger and the minima appear at
a lower $x_{\text{L}}$. For example, the energy level $H = 10^0$
corresponds to Fig.~\ref{inverseSquareForceSolution} where $x_{\text{L}}
= 1$. The adaptive Verlet method (see references in \cite{ms-thesis}) is
used for integration. Since there are multiple variants of this method
that differ in operator splitting and timestep adaption, we give
Algorithm \ref{av-algo} to remove ambiguity and to allow easy
verification of the results. It can be expected that the essence of the
result (what is optimal $\beta$) also holds for other similar adaptive
methods. The initial timestep was chosen such that each run reached
$t_{\text{end}}$ in 1000 steps.

\begin{algorithm}
\caption{AdaptiveVerletSimulation($x_0, v_0, t_{\text{end}}, g, \Delta t_0$)}
\label{av-algo}
\begin{algorithmic}[1]
\STATE $\lambda \leftarrow g(x_0)$
\STATE $\Delta \tau \leftarrow \Delta t_0 / \lambda$
\STATE $(x, v, t) \leftarrow (x_0, v_0, 0)$
\WHILE{ $t < t_{\text{end}}$}
\STATE $v \leftarrow v + \Delta \tau \lambda / 2 \cdot 1/x^2$
\STATE $x \leftarrow x + \Delta \tau \lambda / 2 \cdot v$
\STATE $t \leftarrow t + \Delta \tau \lambda / 2$
\STATE $\lambda \leftarrow t + 1/(2/g(x)-1/\lambda)$
\STATE $x \leftarrow x + \Delta \tau \lambda / 2 \cdot v$
\STATE $v \leftarrow v + \Delta \tau \lambda / 2 \cdot 1/x^2$
\STATE $t \leftarrow t + \Delta \tau \lambda / 2$
\ENDWHILE
\end{algorithmic}
\end{algorithm}

\begin{figure}[htb]
  \includegraphics{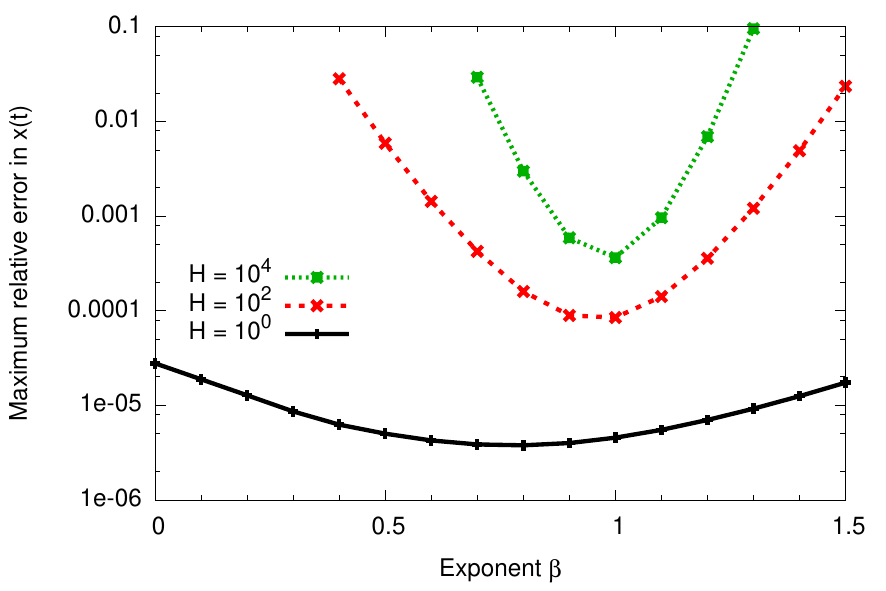}
  \caption{Error of adaptive time step integration depending on exponent
    choice in Sundman transformation. Missong points in the figure indicate that
    the respective simulation runs broke down, e.g.\ because of negative $x$
    values produced during integration.}
  \label{inverseSquareForceErrorVsExponent}
\end{figure}

Figure \ref{inverseSquareForceErrorVsExponent} shows that exponent
$\beta = 1$ is a good choice, where the difference to other exponents gets
pronounced at higher energy levels. We also experimented with other variants of the adaptive
Verlet method and a symplectic adaptive method (see \cite[p. 23]{ms-thesis}). They all showed 
similar behaviour. 


We see a discrepancy compared to what \cite{1dproto} considers as optimal in this so-called hard 
soft-wall problem and note that in \cite{1dproto} not only Coulomb potentials but a whole class of $1/x^{\alpha}$ potentials 
are studied. If we insert $\alpha = 1$ for our case, the choice $\beta = 3/2$ is proposed.
Their findings are based on a scale invariance argument and they verify the result with experiments
 for larger $\alpha$ than in our case. It could be that for the specific case $\alpha = 1$, the general 
 result is too approximative. In our opinion, authors previously using $\beta = 3/2$ should review 
 this choice. 


\section{Applications}
\label{sec:applications}
The presented algorithms were implemented and tested in the \opal\ accelerator simulation framework \cite{opal}. In the first
experiment, we compare AMTS to MTS, and in the second experiment, we compare MTS to the existing
\bb\ integrator. All simulations were performed with small resolution (few thousand particles and small PIC mesh resolution) on a laptop computer.
But it can be expected that the results carry over to large-scale simulations, where of course the absolute gain in computation time is higher.

\subsection{Photoinjector}
\begin{figure}[htb]
\includegraphics{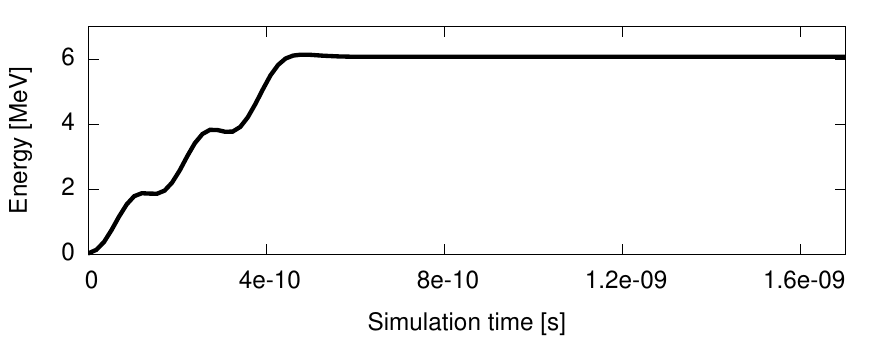}
\caption{Mean particle energy in the photoinjector scenario.}
\label{photoinjectorEnergy}
\end{figure}

\begin{figure}[htb]
\includegraphics{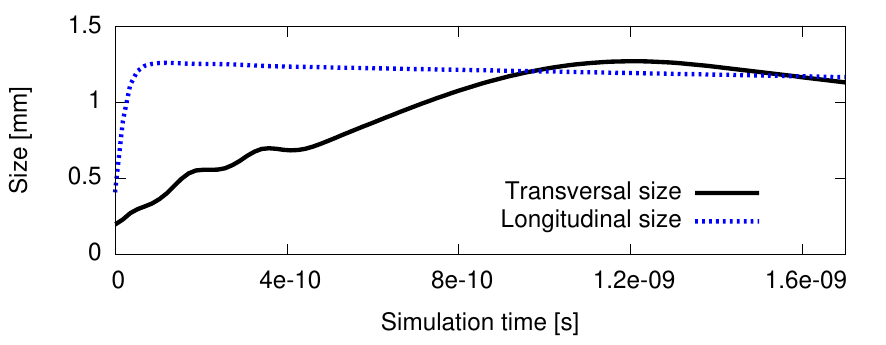}
\caption{Transversal and longitudinal root-mean-square size of the beam for the photoinjector.}
\label{photoinjectorBeamsize}
\end{figure}

We simulate a beam for the first half meter in a photoinjector. This is an example in which space
charge forces are important because of the low beam energy, and where
the magntiude of the space charge forces changes considerably during the
simulation. See Fig.~\ref{photoinjectorEnergy} for the energy curve, and
Fig.~\ref{photoinjectorBeamsize} for the development of the beam size.

\begin{figure}[h]
  \includegraphics{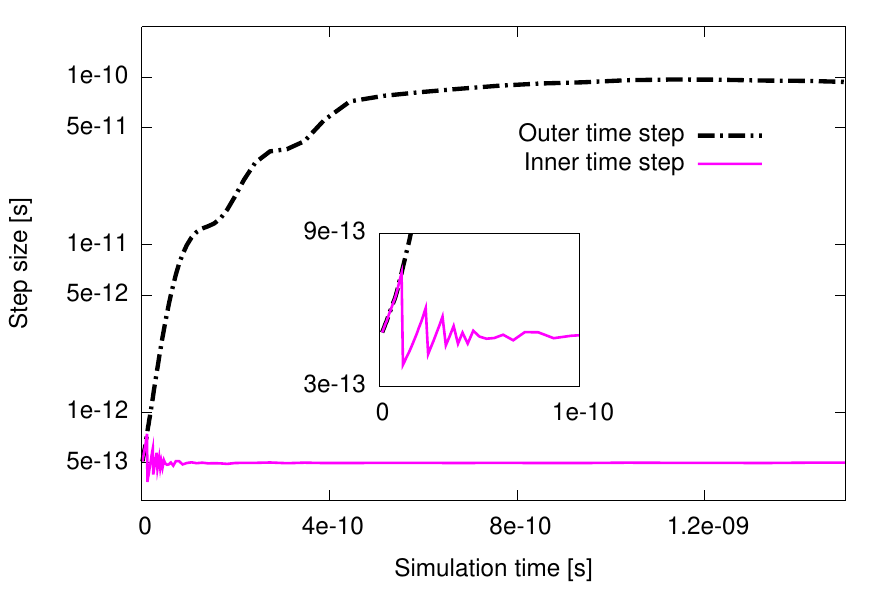}
  \caption{Time step choice of an AMTS run in the photoinjector
    simulation. Because the method requires an integer
    number of substeps per outer step, the inner time step slightly fluctuates around
    the desired value.}
  \label{photoinjectorTimestep}
\end{figure}

We show that the adaptive step size variant of the
multiple-time-stepping integrator (AMTS) is more efficient than the
constant MTS integrator. To
allow a comparison of the different integration strategies with respect
to space charge effects, we want to treat external field integration as
much as possible equally among all runs. In MTS the inner time step is fixed
to $\Delta t_{\text{inner}} = 5 \cdot 10^{-13}$, and the outer step size is a multiple of it. 
In AMTS, as the outer time step
increases, more substeps are used to maintain an inner time step around
$\Delta t_{\text{inner}}$. See Fig.~\ref{photoinjectorTimestep} for the time
step choice of an adaptive run. The
strong variation in the space charge forces leads to a change in the
(outer) time step of more than a factor 100.

\begin{figure}[htb]
  \includegraphics{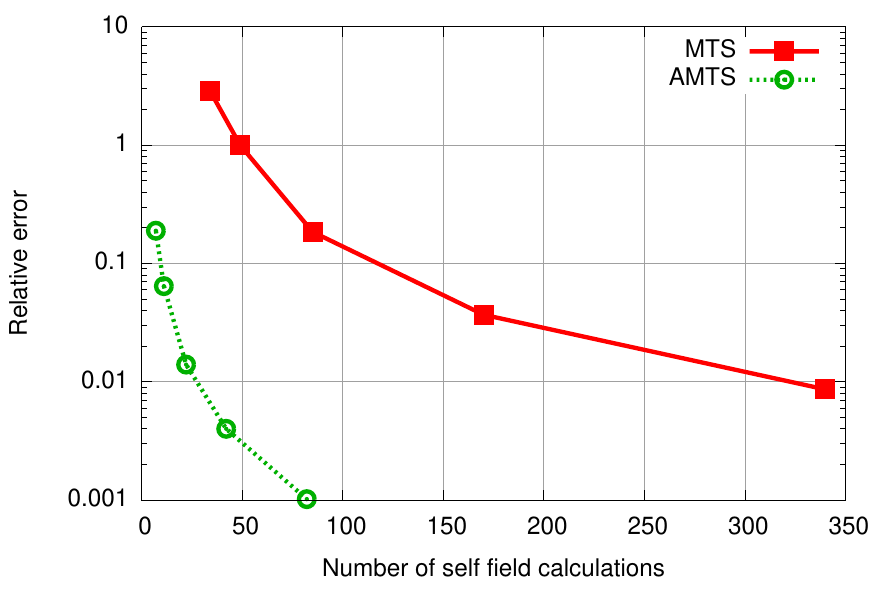}
  \caption{Time integration error in the transverse root-mean-square emittance for the
    photoinjector simulation. Errors are with respect to a reference solution obtained with MTS using 3400 self field calculations.}
  \label{photoinjectorErrorVsWork}
\end{figure}

In Fig.~\ref{photoinjectorErrorVsWork} we give a comparison of the error made
in transverse root-mean-square emittance (a measure of the phase space volume) for different amount of work spent on self
field calculations. The errors are with respect to a reference solution obtained using the MTS integrator with $m = 1$ and a step size of $5\cdot10^{-13}$. AMTS
reaches a given error with drastically fewer self field calculations. 

\begin{figure}[htb]
  \includegraphics{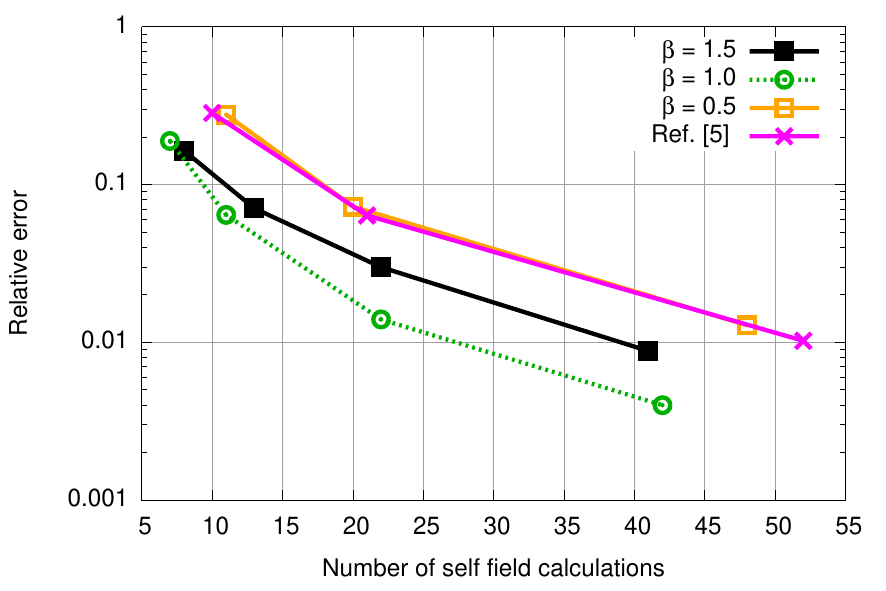}
  \caption{Comparison of different choices of the $g$-function for the AMTS integrator.}
  \label{photoinjectorCompareG}
\end{figure}

In Fig.~\ref{photoinjectorCompareG} we compare different choices of
the $g$-function, which determines how the variable time step is chosen.
Although all choices are far better than a constant step, we see that
our choice with $\beta = 1.0$, derived with the 1D model, gives
slightly better results than the other considered possibilities. It is also
better than the choice from experiments reported in \cite{ms-thesis}, which
used a much simpler adaptation criterion, namely the beam size.

Note that the actual time spent on these simulations is dominated
by the external field integration, because the inner time step is
kept very small.  The reason why we choose the inner time step
unpractically small is that the work spent for external field
integration should be kept constant among the runs, such that
the comparison using the number of self field calculations makes sense. For
practical simulations, the inner time step can be
chosen larger hence for the same accuracy, AMTS yields
a significant reduction in time-to-solution compared to the other integrators.

\subsection{Cyclotron}
The simulation library \opal\ \cite{opal} has an own component for the simulation
of cyclotrons. This component was recently enhanced with support for
neighboring bunch effects \cite{OPAL-CYCL}, increasing significantly the number of space charge calculations in a simulation. 
For this comparison, we simulate the first 10 turns of the PSI 590 MeV ring cyclotron, with
an initial particle energy of 72 MeV. First, we had to recognize that a variable
step size (AMTS) brings no visible benefit over the plain MTS algorithm.
This is understandable as the variation in space charge forces is far smaller than
in the photoinjector case. Also, the influence of space charge is smaller because the
beam energy is higher. Therefore, we use this case to report about the difference
between the original integrator and the MTS variant.

We now want to observe how close a run with reduced space
charge solve frequency (MTS with $m > 1$) comes to the original solution, and how much time
we save. As measure, we take again transversal
RMS emittance. Relative errors are calculated at
five points per turn, and then the maximum of these errors
counts as overall error. The timings are not meant to be a precise profiling, 
but should give a hint how much time is roughly saved.
In general, the savings depend mainly on the fraction of space charge solve cost to 
overall cost of the simulation, beside further details in the implementation. See
Table~\ref{tab:cyclotronMTS} for the results.  Space charge has
certainly an influence in this scenario, as the error of the solution without simulating
space charge is not small. The surprising fact is that a
drastically reduced space charge solve frequency is still very accurate,
but saves a lot of time.

\begin{table}[h]
\begin{center}
\footnotesize  
\begin{tabular}{lll}
\toprule
SC solve frequency & Error in \% & CPU time (seconds) \\
\midrule
$m=1$ & 0 & 524 \\
$m=2$ & $3.35 \cdot 10^{-6}$ & 357 \\
$m=4$ & $1.52 \cdot 10^{-5}$ & 273 \\
$m=10$ & $1.15 \cdot 10^{-4}$ & 222 \\
$m=20$ & $5.13 \cdot 10^{-4}$ & 207 \\
$m=100$ & $8.89 \cdot 10^{-2}$ & 193 \\
No space charge & 12.0 & 190 \\
\bottomrule
\end{tabular}
\end{center}
  \caption{Maximal relative error in transversal RMS emittance and timings
    of MTS integrator in the ring cyclotron simulation. Increasing the
    MTS substep number $m$ (see \eqref{eq:MTS}) reduces the number of self field
    computations, thus saving time while introducing some deviation to
    the solution with full solve frequency ($m = 1$). The solution with no
    space charge effects modeled is given as comparison.}
      \label{tab:cyclotronMTS}
\end{table}
\normalsize

The existing \bb\ implementation in \opal\ (without MTS extension) had
an option to calculate the space charge forces only every $n$-th step,
and then reusing the old forces in the next $n-1$ time steps. This was a
first step towards reducing computation cost. In this context it is
interesting to see how this first approach compares to the new MTS
method. In Table~\ref{tab:cyclotronKeepConst} we give errors and timings
for this older approach. Again, as the timings depend on implementation
details, only the big picture is important here. It is apparent that this, without prejustice
seemingly reasonable scheme (reuse something that changes slowly),
builds up a large error even with low values of $n$, and was therefore
rarely used.  It seems that reusing of old forces introduces an
asymmetry which hurts the accuracy, and a proper MTS implementation
should be used instead.

\begin{table}[h]
\begin{center}
\footnotesize
\begin{tabular}{lll}
\toprule
SC solve frequency & Error in \% & CPU time (seconds) \\
\midrule
$n=1$ & 0 & 473 \\
$n=2$ & $1.01 \cdot 10^{-1}$ & 344 \\
$n=4$ & $2.62 \cdot 10^{-1}$ & 281 \\
$n=10$ & $1.45$ & 242 \\
$n=20$ & $5.46$ & 230 \\
$n=100$ & $15.4$ & 220 \\
No space charge & 12.0 & 197 \\
\bottomrule
\end{tabular}
\end{center}
  \caption{Maximal relative error in transversal RMS emittance and timings
    of modified \bb\ integrator in the ring cyclotron
    simulation. Calculating the self field only 
    every $n$-th step reduces the number of self field computations,
    thus saving time while introducing some deviation to the solution
    with full solve frequency ($n = 1$).  
    The solution with no space charge effects modeled is given as
    comparison.}
   \label{tab:cyclotronKeepConst}
\end{table}
\normalsize


\section{Conclusions}
\label{sec:conclusions}

We presented two time integration schemes that enhance the standard \bb\
algotitm by adapting the frequency of self field calculations.  The
usability was demonstrated within the \opal\ particle accelerator
framework, but the methods could be easily applied to similar problems.

The multiple time stepping (MTS) extension allows to compute the self
fields less frequently, by a factor which has to be defined
beforehand.  For many scenarios where space charge has a visible
contribution but is not dominant, this method can save considerable
computation time with only negligibly changing the solution.  While the
multiple time stepping strategy is not new, we have not seen it applied
in this context.  Initially intended to be an intermediate algorithm only to
derive the variable step size variant, MTS itself turned out to be of
practical relevance, as its implementation incurs hardly any overhead
while the performance gain is substantial.

The adaptive multiple time stepping (AMTS) algorithm introduces a
variable step size integrator.  Variable step sizes are most beneficial
over MTS in space charge dominated (low energy) simulations.  If space
charge forces change considerably, like in gun simulations, AMTS
shows its strengths.  While previous work indicated that variable step
sizes can be useful in such cases, this work gives further insight on
how the step size should be adapted for good results.  The foundation on
MTS allows to have two individual time steps, an outer one that can be
adapted to self field situation, and an inner one that can be kept small
for external fields. The implementation is more complicated than for
MTS.  However, we have shown that important problems exist for which
this additional effort pays off.


\section*{Acknowledgements}
We are grateful to A.~Fallahi for discussions related to the analysis
of the single particle model problem, to S.~Wei for performing first
practical tests with the MTS method, and to J.~Conrad for financial support.

\section*{References}
\bibliographystyle{model1-num-names}
\bibliography{biblio}

\begin{thebibliography}{13}
\expandafter\ifx\csname natexlab\endcsname\relax\def\natexlab#1{#1}\fi
\providecommand{\bibinfo}[2]{#2}
\ifx\xfnm\relax \def\xfnm[#1]{\unskip,\space#1}\fi
\bibitem[{Adelmann et~al.(2010)Adelmann, Kraus, and et.al}]{opal}
\bibinfo{author}{A.~Adelmann}, \bibinfo{author}{C.~Kraus},
  \bibinfo{author}{Y.~I. et.al}, \bibinfo{title}{{The OPAL (Object Oriented
  Parallel Accelerator Library) Framework}}, \bibinfo{type}{Technical Report}
  \bibinfo{number}{PSI-PR-08-02}, Paul Scherrer Institut,
  \bibinfo{year}{2008-2010}.
  \bibinfo{note}{\url{http://amas.web.psi.ch/docs/opal/opal_user_guide.pdf}}.
\bibitem[{Agarwalla et~al.(2011)Agarwalla, Conrad, and
  Shaevitz}]{AgarwallaConradShaevitz}
\bibinfo{author}{S.~K. Agarwalla}, \bibinfo{author}{J.~M. Conrad},
  \bibinfo{author}{M.~Shaevitz},
\newblock \bibinfo{title}{{Short-baseline Neutrino Oscillation Waves in
  Ultra-large Liquid Scintillator Detectors}},
\newblock \bibinfo{journal}{arXiv} \bibinfo{volume}{1105.4984, hep-ph}
  (\bibinfo{year}{2011}).
\bibitem[{Abs et~al.(2012)Abs, Adelmann et~al.}]{daedalus}
\bibinfo{author}{M.~Abs}, \bibinfo{author}{A.~Adelmann}, et~al.,
\newblock \bibinfo{title}{{Multimegawatt DAE$\delta$ALUS Cyclotrons for
  Neutrino Physics}},
\newblock \bibinfo{journal}{arXiv} \bibinfo{volume}{1207.4895, physics.acc-ph}
  (\bibinfo{year}{2012}).
\bibitem[{Bungau et~al.(2012)Bungau, Adelmann et~al.}]{isodar}
\bibinfo{author}{A.~Bungau}, \bibinfo{author}{A.~Adelmann}, et~al.,
\newblock \bibinfo{title}{{An Electron Antineutrino Disappearance Search Using
  High-Rate 8Li Production and Decay}},
\newblock \bibinfo{journal}{Phys.Rev.Lett.} \bibinfo{volume}{109}
  (\bibinfo{year}{2012}) \bibinfo{pages}{141802}.
\bibitem[{Toggweiler(2011)}]{ms-thesis}
\bibinfo{author}{M.~Toggweiler}, \bibinfo{title}{An adaptive time integration
  method for more efficient simulation of particle accelerators}, Master's
  thesis, ETH Zurich, \bibinfo{year}{2011}.
  \bibinfo{note}{\url{http://e-collection.library.ethz.ch/list/author?author=Toggweiler\%2C+Matthias}}.
\bibitem[{Gear(1971)}]{gear71}
\bibinfo{author}{C.~W. Gear}, \bibinfo{title}{Numerical initial value problems
  in ordinary differential equations}, \bibinfo{publisher}{Prentice--Hall},
  \bibinfo{address}{Englewood Cliffs, N.J}, \bibinfo{year}{1971}.
\bibitem[{Landau and Lifshitz(1984)}]{lali:84}
\bibinfo{author}{L.~D. Landau}, \bibinfo{author}{E.~M. Lifshitz},
  \bibinfo{title}{Electrodynamics of Continuous Media},
  \bibinfo{publisher}{Pergamon}, \bibinfo{address}{Oxford},
  \bibinfo{edition}{2nd} edition, \bibinfo{year}{1984}.
\bibitem[{Qiang and Ryne(2001)}]{qiry:01}
\bibinfo{author}{J.~Qiang}, \bibinfo{author}{R.~D. Ryne},
\newblock \bibinfo{title}{Parallel {3D} {P}oisson solver for a charged beam in
  a conducting pipe},
\newblock \bibinfo{journal}{Comput. Phys. Commun.} \bibinfo{volume}{138}
  (\bibinfo{year}{2001}) \bibinfo{pages}{18--28}.
\bibitem[{Qiang and Gluckstern(2004)}]{qigl:04}
\bibinfo{author}{J.~Qiang}, \bibinfo{author}{R.~L. Gluckstern},
\newblock \bibinfo{title}{Three-dimensional {P}oisson solver for a charged beam
  with large aspect ratio in a conducting pipe},
\newblock \bibinfo{journal}{Comput. Phys. Commun.} \bibinfo{volume}{160}
  (\bibinfo{year}{2004}) \bibinfo{pages}{120--128}.
\bibitem[{Birdsall and Langdon(1985)}]{plasmaPhysics}
\bibinfo{author}{C.~K. Birdsall}, \bibinfo{author}{A.~B. Langdon},
  \bibinfo{title}{Plasma Physics via Computer Simulation},
  \bibinfo{publisher}{McGraw-Hill}, \bibinfo{year}{1985}.
\bibitem[{Hairer et~al.(2006)Hairer, Lubich, and Wanner}]{geomNumInt}
\bibinfo{author}{E.~Hairer}, \bibinfo{author}{C.~Lubich},
  \bibinfo{author}{G.~Wanner}, \bibinfo{title}{Geometric Numerical Integration:
  Structure-Preserving Algorithms for Ordinary Differential Equations},
  \bibinfo{publisher}{Springer}, \bibinfo{address}{Berlin},
  \bibinfo{year}{2006}.
\bibitem[{Bond and Leimkuhler(1998)}]{1dproto}
\bibinfo{author}{S.~Bond}, \bibinfo{author}{B.~Leimkuhler},
\newblock \bibinfo{title}{Time-transformations for reversible variable stepsize
  integration},
\newblock \bibinfo{journal}{Numer. Algorithms} \bibinfo{volume}{19}
  (\bibinfo{year}{1998}) \bibinfo{pages}{55--71}.
\bibitem[{Yang et~al.(2010)Yang, Adelmann, Humbel, Seidel, and
  Zhang}]{OPAL-CYCL}
\bibinfo{author}{J.~Yang}, \bibinfo{author}{A.~Adelmann},
  \bibinfo{author}{M.~Humbel}, \bibinfo{author}{M.~Seidel},
  \bibinfo{author}{T.~Zhang},
\newblock \bibinfo{title}{{Beam Dynamics in High Intensity Cyclotrons Including
  Neighboring Bunch Effects: Model, Implementation and Application}},
\newblock \bibinfo{journal}{Phys. Rev. ST Accel. Beams} \bibinfo{volume}{13}
  (\bibinfo{year}{2010}) \bibinfo{pages}{064201}.

\end{thebibliography}

\end{document}